\documentclass[12pt]{article}

\usepackage[a4paper,margin=25mm]{geometry}
\usepackage{graphicx}
\usepackage{hyperref}
\usepackage{amsmath}
\usepackage{booktabs}
\usepackage{array}
\usepackage{seqsplit}
\usepackage{setspace}

\newcolumntype{L}[1]{>{\raggedright\arraybackslash}p{#1}}
\newcommand{\botrule}{\bottomrule}
\title{Reduced latent leakage does not reliably predict lower likelihood bias in collider inference}
\author{Tong Pan\\
\small College of Mathematics and Physics, Suqian University, Suqian, Jiangsu, China\\
\small \href{mailto:23211@squ.edu.cn}{23211@squ.edu.cn}\\
\small ORCID: \href{https://orcid.org/0000-0002-4700-1516}{0000-0002-4700-1516}}
\date{}

\begin{document}

\maketitle

\begin{abstract}
Reusable collider representations can be evaluated through downstream discrimination and probes of retained information, but neither quantity directly tests the behaviour of score templates in a profiled likelihood. We test a specific prediction in a controlled two-channel routing protocol: if reduced physics-label readability in a nuisance branch indicates a more inference-robust representation, it should accompany a smaller profiled signal-strength bias under fixed unmodelled shifts. In a public Compact Muon Solenoid $H\rightarrow ZZ\rightarrow4\ell$ workflow, a downstream split of fixed EveNet embeddings preserves signal/background area under the receiver operating characteristic curve ($0.9894\pm0.0004$) while reducing nuisance-branch physics readability from $0.961\pm0.013$ to $0.593\pm0.030$. Probe-sensitivity and effective-rank controls exclude a failed readout and branch collapse. In a separate top quark jet-tagging workflow, the leakage reduction recurs with preserved task performance. Across two development event shards, however, its Spearman association with maximum absolute profiled bias is $0.036$, and three of six material leakage-improving transitions do not reduce that bias. A one-shot preregistered confirmation on an independently accessed shard produces material leakage reductions in all three paired seeds, while the maximum absolute bias increases in two. Thus, within the tested protocol, latent readability is a useful routing diagnostic but not a likelihood-robustness certificate. The result supports a practical validation rule: claims about inference robustness require a prespecified likelihood-facing stress test and held-out confirmation.
\end{abstract}

\noindent\textbf{Keywords:} high-energy physics; representation learning; systematic uncertainty; profile likelihood; robustness audit

\section{Introduction}

Machine learning in high energy physics (HEP) is moving beyond isolated classifiers toward reusable representations of particles, jets and collision events~\cite{birk2024omnijet,bardhan2025hepjepa,ho2024pretrained,hsu2026evenet,hallin2025foundation}. Reuse can lower the cost of adapting a model to a new task, but it also separates representation development from the statistical model used in a specific analysis. A representation may transfer well as a classifier while its score templates respond unfavourably to a systematic shift. Discrimination performance alone cannot resolve that question.

Existing work addresses nuisance information at several points in this chain. Domain-adversarial and shared-private methods shape which information a representation retains~\cite{ganin2016domain,bousmalis2016domain}. Adversarial pivoting seeks nuisance-insensitive predictions, while Inference-Aware Neural Optimisation (INFERNO), uncertainty-aware learning and systematics-aware neural training align optimisation more directly with statistical inference~\cite{louppe2017pivot,decastro2019inferno,ghosh2021uncertainty,birk2025sannt}. These approaches motivate two separate validation questions. The first asks whether nuisance or task information remains readable from a representation. The second asks how the resulting score templates behave after nuisance profiling. Success on the first question does not logically answer the second because the profiled endpoint depends on bin-wise signal and background responses, nuisance constraints and shifts omitted from the fitted model.

The distinction is related to an earlier warning about decorrelation. Ghosh and Nachman showed that decorrelating selected generator variations can reduce an estimated theory uncertainty without reducing disagreement with an independent generator~\cite{ghosh2022caution}. Their study concerns a theory-uncertainty proxy, not nuisance-branch readability or profiled signal-strength bias. We test the corresponding prediction at the representation-audit layer. If lower physics-label readability in a nuisance branch is evidence that a two-channel representation is safer for inference, then paired reductions in that readability should be accompanied by reductions in a frozen likelihood-facing bias metric. A consistent failure of this directional prediction would show that the latent diagnostic is insufficient for certification, while leaving its value as a representation diagnostic intact.

We evaluate this prediction in a controlled two-channel routing model. The nuisance channel should carry domain information without becoming a hidden second task classifier, so lower physics-label readability is a meaningful design objective. The endpoint is the maximum absolute profiled signal-strength bias, $B_{\max}$, under shifted templates that are not included in the fitted nuisance model. The study therefore does not compare two arbitrary metrics. It asks whether improvement in the model's own routing objective predicts improvement at the downstream statistical endpoint for which robustness matters.

Figure~\ref{fig:overview_schematic} summarizes the design. A public $H\rightarrow ZZ\rightarrow4\ell$ (H4l) workflow first tests whether a downstream split of fixed EveNet embeddings can preserve discrimination while reducing nuisance-branch physics readability. Probe and rank controls check for weak readout and branch collapse. A separate top quark jet-tagging (TopTag) workflow then supplies systematic-domain score templates, a development-shard likelihood audit and a one-shot preregistered confirmation on an independently accessed event shard. ``Cross-workflow consistency'' refers only to recurrence of the routing effect in H4l and TopTag. The likelihood mismatch and its held-out confirmation both belong to the TopTag workflow; they are not a new detector analysis or a cross-method replication.

The central empirical result is that a large, reproducible reduction in nuisance-branch physics readability does not reliably predict a smaller profiled-likelihood bias within this protocol. H4l provides enabling evidence that the routing intervention is material, preserves task performance and survives probe and collapse controls. TopTag supplies the paired proxy-to-endpoint test, including the preregistered shard confirmation. The practical output is a minimum validation sequence that keeps representation diagnostics, likelihood-facing stress tests and held-out confirmation as distinct evidentiary layers. The claim is deliberately local to the tested routing protocol, shifts and likelihood construction; it does not assert that every latent metric or nuisance-aware method fails.

\section{Methods}
\label{sec:methods}

\subsection{Study design}

\begin{figure*}[t]
\centering
\includegraphics[width=0.98\textwidth]{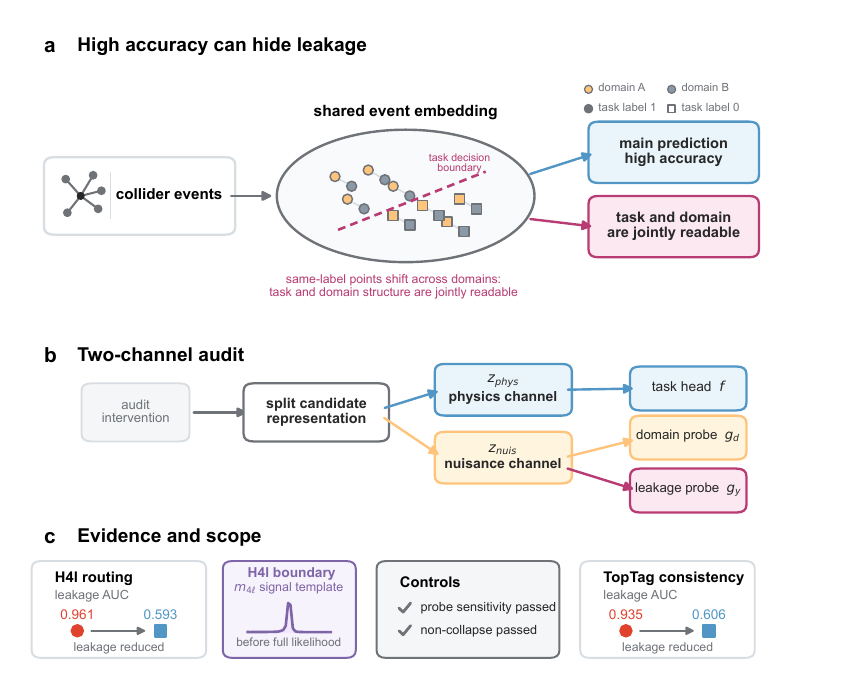}
\caption{\textbf{Two-stage representation audit.}
\textbf{a,} High task accuracy can coexist with jointly readable task and domain structure. Here a domain denotes the data-generation or observation conditions, such as different simulation, detector-response or reconstruction settings. Circles and squares denote the two task labels, while yellow and grey denote domains A and B. Thin connectors pair schematic instances with the same task label across domains, showing a domain response without requiring the points to cross the task boundary. The magenta dashed line is the task decision boundary, not a domain boundary. \textbf{b,} The downstream audit intervention used in this study separates a physics channel, $z_{\mathrm{phys}}$, from a nuisance channel, $z_{\mathrm{nuis}}$; EveNet itself supplies one fixed embedding rather than these two channels. The task head reads $z_{\mathrm{phys}}$, whereas post hoc domain and physics-leakage probes audit $z_{\mathrm{nuis}}$ after training. \textbf{c,} H4l and TopTag establish that the split materially reduces nuisance-channel leakage without task collapse. The template cues mark the required second stage: the latent change must be checked against an analysis-facing likelihood outcome rather than treated as a certificate by itself. Red circles denote the shared baseline, blue squares the split candidate, yellow the nuisance/domain path, magenta the leakage cues, grey the controls and purple the analysis boundary.}
\label{fig:overview_schematic}
\end{figure*}

The audit is organized around two questions. First, can a downstream intervention materially reduce physics-label readability from a nuisance branch while preserving the intended task and a noncollapsed representation? Second, if that latent proxy improves, does the maximum absolute profiled signal-strength bias also decrease under an unmodelled shift? The first question validates the intervention; the second tests the prespecified proxy-to-endpoint prediction (Fig.~\ref{fig:overview_schematic}a).

Each study compares a shared baseline with a split candidate using the same events, train/validation/test split and probe procedure. The shared baseline supplies one latent representation to all readout heads. The split candidate instead produces a physics channel, $z_{\mathrm{phys}}$, and a nuisance channel, $z_{\mathrm{nuis}}$ (Fig.~\ref{fig:overview_schematic}b). The task head $f$ predicts the physics label from $z_{\mathrm{phys}}$. After representation training is complete, two separate probes read $z_{\mathrm{nuis}}$: the domain probe $g_d$ tests whether domain variation is detectable, and the leakage probe $g_y$ tests whether the physics label remains detectable. These probes diagnose the fixed representation; they are not additional task heads used to make the primary prediction.

The first-stage paired requirement asks the split candidate to preserve task performance through $z_{\mathrm{phys}}$ while reducing physics-label leakage from $z_{\mathrm{nuis}}$. Probe-sensitivity and effective-rank controls address the two main trivial explanations: an unreadable probe and an empty branch. Passing this stage establishes that the latent proxy has changed nontrivially; it does not certify inference robustness. H4l provides the main routing and control evidence, while TopTag first checks cross-workflow consistency and then supplies the frozen likelihood endpoint. Development shards are used to identify the candidate mismatch; a separate preregistered shard tests it without post-access tuning.

The main validation uses a CMS $H\rightarrow ZZ\rightarrow4\ell$ workflow derived from public open data because it places the audit in a recognizable HEP analysis topology that can be reconstructed from public records. The secondary validation uses a TopTag systematic workflow to test the same audit logic in a particle-level tagging benchmark; rejection at fixed efficiency provides an additional operating-point metric. The H4l study uses synthetic visible-domain controls rather than official CMS systematic uncertainties. The TopTag study uses the nominal, energy-scale, cluster-energy-resolution, cluster-position and bias domains supplied with the open dataset.

\subsection{Branch routing objective and metrics}

The branch routing objective combines a physics task loss with auxiliary domain routing and branch separation terms. Given event features $x$, the representation encoder returns a physics channel, $z_{\mathrm{phys}}$, and a nuisance channel, $z_{\mathrm{nuis}}$. A physics head, denoted by $f$, predicts the signal/background label, $y_{\mathrm{phys}}$, from $z_{\mathrm{phys}}$. Post hoc probe heads, denoted by $g$, are trained after the representation is fixed to test how much physics and domain information remain readable from each channel.

For a split model, the implemented training objective has the form
\begin{align}
\mathcal{L}_{\mathrm{split}}={}&
\mathcal{L}_{\mathrm{phys}}+
\lambda_d\mathcal{L}_{\mathrm{domain}}+
\lambda_a\mathcal{L}_{\mathrm{adv}} \notag\\
&+\frac{\lambda_o}{p q}
\left\|\widetilde{Z}_{\mathrm{phys}}^{\mathsf T}
\widetilde{Z}_{\mathrm{nuis}}\right\|_F^2 .
\end{align}
where $\mathcal{L}_{\mathrm{phys}}$ is class-weighted binary cross entropy, $\mathcal{L}_{\mathrm{domain}}$ is cross entropy from $z_{\mathrm{nuis}}$, and $\mathcal{L}_{\mathrm{adv}}$ is domain cross entropy behind a gradient-reversal layer on $z_{\mathrm{phys}}$. The columns of $\widetilde{Z}$ are batch-centred and unit-normalised, and $p$ and $q$ are the two branch dimensions. Terms absent from a candidate are assigned zero weight; the shared baseline uses only $\mathcal{L}_{\mathrm{phys}}$. Exact H4l and TopTag weights, dimensions and optimisation settings are listed in Appendix~\ref{app:training_settings}. The post hoc probes do not enter this objective.

The primary leakage metric is the probe AUC for predicting the physics label from $z_{\mathrm{nuis}}$,
\[
L_{\mathrm{phys\ from\ nuis}} = \mathrm{AUC}\{g(z_{\mathrm{nuis}}), y_{\mathrm{phys}}\}.
\]
Here $\mathrm{AUC}\{\hat{y}, y\}$ denotes the ranking metric obtained from predictions $\hat{y}$ and labels $y$. Lower values indicate that information useful for physics discrimination is less readable from the nuisance channel. This is a leakage metric, not proof of causal removal. We report it together with the primary physics AUC from $z_{\mathrm{phys}}$,
\[
A_{\mathrm{phys}} = \mathrm{AUC}\{f(z_{\mathrm{phys}}), y_{\mathrm{phys}}\},
\]
so that leakage reduction cannot be overclaimed if the useful physics task collapses.

Domain readability is measured analogously with probe accuracy for the controlled domain label $d$. In the H4l workflow the domains are synthetic visible domain variants used for stress testing and are therefore interpreted as controlled proxies rather than CMS nuisance parameters. In the TopTag workflow the domains are nominal and systematic variation samples provided by the open dataset. We also report the maximum score drift across domains,
\[
\Delta_{\mathrm{score}}^{\max} =
\max_d \left| \mathrm{E}[s(x)\mid d] - \mathrm{E}[s(x)\mid d_{\mathrm{nom}}] \right|,
\]
where $s(x)$ is the physics score for event features $x$, $d_{\mathrm{nom}}$ is the nominal domain and $\mathrm{E}[\cdot\mid d]$ is the conditional mean over events in domain $d$. This quantity measures domain stability at the score level and complements the branch leakage probes.
In the Results, these quantities are reported by their reader-facing names: $z_{\mathrm{nuis}}\!\to$physics AUC, physics AUC and maximum score drift, respectively. Domain-probe accuracy is reported separately as the domain-readability check.

\subsection{H4l open data workflow}

The H4l workflow starts from reduced CMS open data records for the $H\rightarrow ZZ\rightarrow4\ell$ outreach analysis and associated simulation samples~\cite{cms_h4l_opendata}. The signal topology is a Higgs decay through two $Z$ bosons into four isolated leptons, for which the four lepton invariant mass $m_{4\ell}$ is the standard analysis anchor against the continuum $ZZ$ background. Events are reconstructed into four lepton candidates and represented as tensors with shape $[N,4,10]$, where $N$ is the number of candidates. The four rows correspond to the selected leptons. The ten features for each lepton are $p_T$, $\eta$, $\phi$, mass, charge, relative isolation, transverse and longitudinal impact parameter significances, and muon/electron indicators. The formal tensor export contains 218618 selected candidates, with 26706 signal and 191912 $ZZ$ background candidates. Five event level conditions accompany each candidate.

The four lepton tensor is processed with EveNet as a fixed feature extractor to obtain a reusable event embedding. We use EveNet as a public representation interface for the reliability protocol, with measurement interpretation left to the standard H4l observables and likelihood machinery. A feasibility check on 218618 embeddings confirmed finite embeddings for all events and gave an embedding probe AUC of 0.703, establishing that the fixed EveNet representation carries nontrivial H4l information before branch routing.

EveNet supplies a single 256-dimensional embedding in this workflow. It does not supply the physics and nuisance branches studied here. Those branches are introduced by a downstream network trained on the frozen EveNet output, so conclusions about their routing behaviour apply to the intervention evaluated in this study rather than to the native EveNet architecture.

The split branch model is trained and evaluated on mixed visible domains. These visible domain shifts are synthetic controls that alter the visible event representation used by the protocol while keeping labels and event weights fixed across visible domain copies. They are used to test whether branch routing can preserve the physics task while reducing physics leakage from the nuisance channel under matched visible domain stress.

The main H4l result across seeds uses seeds 41, 42 and 43 with identical event splits, fixed EveNet embeddings and probe settings after training. The selected split variant is the best tradeoff declared before the main comparison from the small H4l grid: it keeps the physics head on $z_{\mathrm{phys}}$, routes domain pressure through $z_{\mathrm{nuis}}$, and adds an orthogonality penalty to discourage the two branches from carrying the same information. Physics and domain probes are trained after the representation model is fixed. They are diagnostic readouts only; their gradients are not used to update the fixed EveNet features or to change the reported branch model.

\subsection{Controls and boundary checks}

Several controls are included to avoid mistaking a weak probe or a collapsed representation for disentanglement. A control with shuffled domains measures the domain probe after domain labels are randomly permuted. A control with an injected one-hot label appends a direct domain code to verify that the probe can recover domain information when it is present. A teacher embedding control tests whether the fixed EveNet representation itself carries domain readable information. Effective rank checks are computed for $z_{\mathrm{phys}}$ and $z_{\mathrm{nuis}}$ to test whether either branch collapses to a nearly degenerate representation.

We also compare learned scores with H4l checks that face the analysis problem more directly. A mass template comparison uses the four lepton invariant mass, $m_{4\ell}$, as the conventional physics baseline. A label-free observed data overlay compares observed and simulated $m_{4\ell}$ shapes in an exploratory sanity check normalised in the sidebands. These checks set the analysis boundary: they compare the learned score with the established $m_{4\ell}$ reference without constructing a full CMS likelihood, object correction chain or nuisance model.

\subsection{TopTag systematic workflow}

The TopTag study uses the ATLAS Top Tagging Open Data Set with systematic variations~\cite{atlas_toptag80030}. The physics task is boosted hadronic top tagging: top jets contain characteristic substructure with multiple prongs from $t\rightarrow bW\rightarrow bq\bar{q}'$, while quantum chromodynamics (QCD) background jets are dominated by ordinary quark and gluon radiation patterns. We use the first balanced shards from nominal, energy scale up/down, cluster energy resolution, cluster position and bias domains, with 100000 events per domain and 600000 events per run. A small DeepSets-style constituent encoder is used to avoid relying on the EveNet feature extractor, which is specific to H4l. The comparison keeps the same protocol logic: preserve top tagging AUC from $z_{\mathrm{phys}}$ while reducing physics leakage from $z_{\mathrm{nuis}}$ and measuring domain readability.

A separate TopTag reference dataset run calibrates model capacity on the standard reference dataset for top quark tagging~\cite{toptag_reference}. This calibration is not a systematic study; it checks that the small constituent encoder is a competent tagger before it is used in the systematic study.

\subsection{Metric-to-likelihood audit and held-out event-shard confirmation}

The likelihood-facing TopTag benchmark uses nominal score templates for top signal and quantum chromodynamics background and evaluates shifted pseudo data with a binned \texttt{pyhf}/HistFactory-style likelihood. The fitted model includes the designated energy-scale morphing nuisance with a Gaussian constraint. The unmodelled-shift rows instead generate pseudo data from a different systematic domain that is absent from the fitted nuisance model. For each model and seed, the analysis-facing endpoint is the maximum absolute fitted signal-strength bias across these unmodelled shifts, denoted $B_{\max}=\max |\hat{\mu}-1|$. This is a controlled template stress test, not an ATLAS measurement or a calibrated experimental uncertainty model.

The development audit freezes the exact shared, balanced split and frozen-residual checkpoints already obtained on event shards 000 and 001. It evaluates three independently initialized post hoc probes per checkpoint, using the same event partition, and compares their mean nuisance-branch physics AUC with $B_{\max}$. The 18 model rows reuse shards and seeds and are therefore not treated as independent experimental units; their Spearman rank association is a descriptive development statistic. The prespecified material-change threshold is a leakage-AUC decrease of at least 0.05. This is an operational effect-size gate, chosen to exclude marginal probe fluctuations rather than a universal calibration of scientific relevance. We count paired transitions for which leakage improves materially but $B_{\max}$ does not decrease. The development audit cannot by itself provide held-out confirmation because the two shards had already informed the analysis sequence.

Before accessing event shard002, we froze a separate confirmation configuration, the shared-to-balanced comparison, seeds 41--43, three probe seeds, the 20-bin template construction, the likelihood settings and the decision rules. Nominal signal-plus-background bin sums form Asimov observations; energy-scale up/down templates define the fitted \texttt{histosys}, with $0\leq\mu\leq5$ and $-5\leq\theta_{\mathrm{ES}}\leq5$. Cluster-resolution, cluster-position and bias templates are substituted as pseudo data but are absent from the fitted nuisance model, and $B_{\max}$ takes the maximum across those three rows. Empty template bins are floored at $10^{-6}$. Confirmation required at least two material leakage improvements, preservation of physics AUC within 0.01 in every pair and a cleaner-but-not-better fraction of at least one third among material transitions. These are frozen operational gates: the AUC tolerance prevents a task-collapse explanation, while the count and fraction rules require the discordance to recur rather than depend on one seed; they are not claimed as universal thresholds or significance tests. The six shard002 files, their sizes and Adler-32 checksums were also fixed before access. Shard002 was then accessed once; no seed, architecture, probe, binning, nuisance, likelihood, threshold or proxy sweep was permitted after access.

\section{Results}
\label{sec:results}

\subsection{The H4l workflow provides a representation test close to analysis use}

We first verified that the routing protocol could be evaluated in a recognizable public H4l workflow. Preprocessing selected 218618 four lepton candidates; a mass-window counting proxy gave weighted $S/\sqrt{B}=2.39$, and a mass-proximity classifier reached AUC 0.990. EveNet then produced finite 256 dimensional embeddings for every selected candidate, with embedding probe AUC 0.703. These checks verify the intended H4l structure and a nontrivial fixed representation interface; they are workflow inputs rather than evidence for the proxy-to-likelihood claim.

\subsection{Branch routing suppresses physics leakage from the nuisance channel while preserving H4l AUC}

\begin{figure*}[t]
\centering
\includegraphics[width=0.80\textwidth]{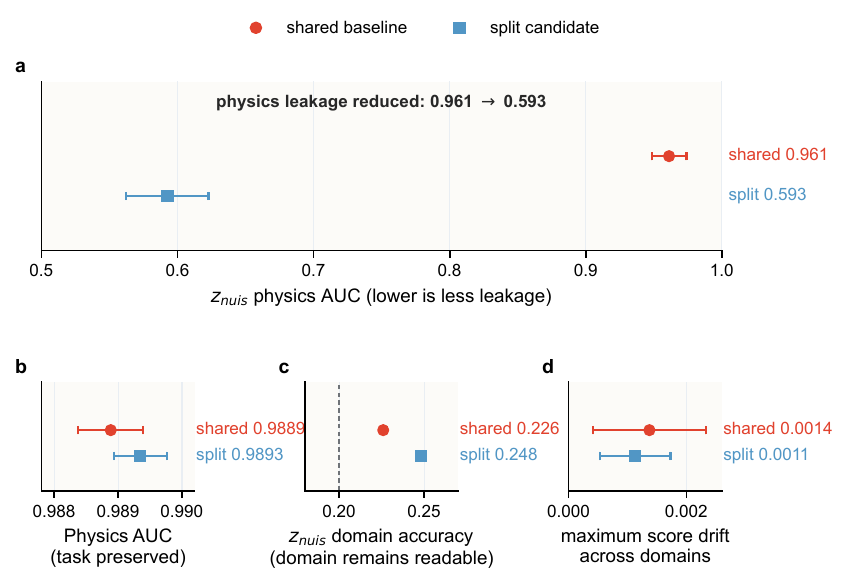}
\caption{\textbf{H4l branch routing result.}
The split candidate preserves H4l physics AUC while reducing physics leakage from the nuisance channel. Red circles denote the shared baseline and blue squares denote the split candidate throughout the figure; markers and horizontal bars show mean $\pm$ standard deviation (SD) over three seeds. \textbf{a,} Physics-label AUC measured from $z_{\mathrm{nuis}}$; lower values mean less physics leakage, and the neutral annotation reports the shared-to-split change. \textbf{b,} Physics AUC from $z_{\mathrm{phys}}$, showing that the task is preserved. \textbf{c,} Domain-probe accuracy from $z_{\mathrm{nuis}}$; the dashed vertical line marks the five-domain random expectation of 0.20. \textbf{d,} Maximum physics-score drift across domains.}
\label{fig:h4l_branch_routing}
\end{figure*}

To test whether branch routing can suppress leakage without sacrificing the task, we trained shared and split models on the same three mixed visible domain seeds (Fig.~\ref{fig:h4l_branch_routing}). The shared baseline and split candidate have nearly identical physics performance from the physics channel. The shared baseline reaches physics AUC $0.9889\pm0.0005$, while the split candidate reaches $0.9894\pm0.0004$. The main difference is where physics information remains readable. The shared baseline has $z_{\mathrm{nuis}}\rightarrow$physics AUC $0.961\pm0.013$, whereas the split candidate reduces this leakage to $0.593\pm0.030$.

Together, these measurements validate the first stage of the audit. Physics discrimination is preserved in $z_{\mathrm{phys}}$, physics readout from $z_{\mathrm{nuis}}$ is strongly reduced, and score drift remains small under the tested visible-domain controls. This is a nontrivial routing result, not yet a likelihood result.

\subsection{Controls rule out trivial probe failure and branch collapse}

\begin{figure*}[t]
\centering
\includegraphics[width=0.80\textwidth]{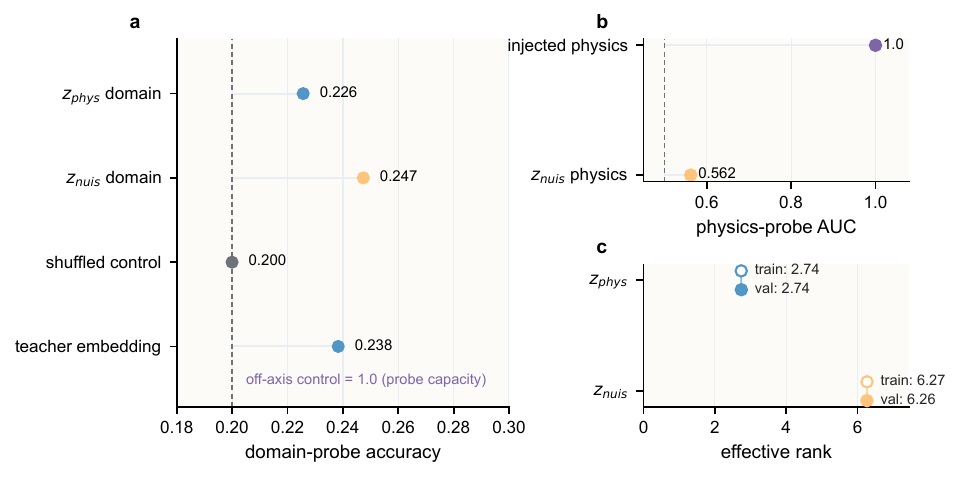}
\caption{\textbf{H4l probe and collapse controls.}
Controls rule out weak probes and empty branches as explanations for the leakage reduction. Blue and yellow denote $z_{\mathrm{phys}}$ and $z_{\mathrm{nuis}}$, grey denotes shuffled or reference controls, and purple denotes injected positive controls. \textbf{a,} Domain-probe accuracy for $z_{\mathrm{phys}}$, $z_{\mathrm{nuis}}$, shuffled labels and the fixed EveNet teacher embedding. The dashed vertical line is the five-domain random expectation of 0.20; the injected domain control reaches 1.0. \textbf{b,} Physics-probe AUC for an injected physics control and for $z_{\mathrm{nuis}}$; the dashed vertical line marks random binary classification at 0.5. \textbf{c,} Effective rank of $z_{\mathrm{phys}}$ and $z_{\mathrm{nuis}}$ on train and validation samples. Open and filled circles denote train and validation values, respectively.}
\label{fig:h4l_controls}
\end{figure*}

To test whether the leakage reduction could be an artifact of a weak probe or a collapsed branch, we ran controls with shuffled labels, injected domains and rank checks (Fig.~\ref{fig:h4l_controls}). In the best H4l variant, physics AUC is 0.990 and maximum score drift across domains is 0.0012. The true $z_{\mathrm{nuis}}$ domain probe is 0.247, while the shuffled domain control drops to 0.200, near the random expectation for five domains. The injected one-hot domain control reaches 1.0, confirming that the probe can recover domain information when it is explicitly present. The teacher embedding domain probe is 0.238, indicating that some domain readable structure is already present in the fixed EveNet representation.

The rank checks for the representation argue against collapse. The effective rank is 2.74 for $z_{\mathrm{phys}}$ and approximately 6.26 for $z_{\mathrm{nuis}}$, with no inspected branch dimensions near zero. Thus the leakage reduction is not explained by a branch becoming numerically empty. The remaining $z_{\mathrm{nuis}}\rightarrow$physics AUC of 0.562 in the control run shows that physics information is reduced in the nuisance branch rather than eliminated.

\subsection{Mass templates and observed data check the H4l workflow}

The lightweight Monte Carlo (MC)/Asimov checks compare learned scores with the standard H4l mass anchor. In these scans, the finely binned $m_{4\ell}$ template remains stronger than templates built from the shared or split learned scores, placing the learned scores in a diagnostic role for the representation study.

The label-free observed-data overlay is retained as an exploratory workflow sanity check (Appendix Fig.~\ref{fig:h4l_boundary}). Because the 267 selected events have not been deduplicated across DoubleMu and DoubleElectron streams, the sideband-normalized comparison is not used to support the representation or likelihood claims; detailed counts and shape checks remain in the Appendix.

\subsection{TopTag cross-workflow consistency}

\begin{figure*}[t]
\centering
\includegraphics[width=0.80\textwidth]{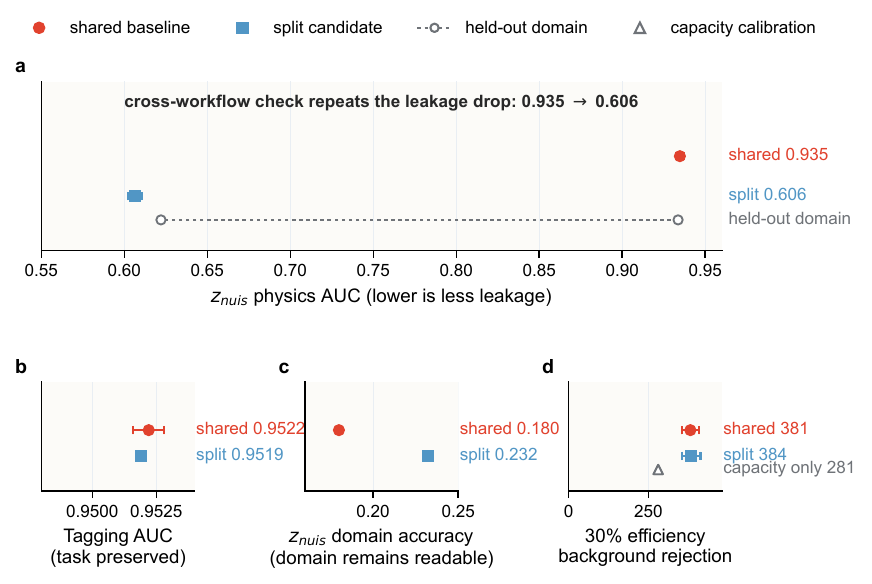}
\caption{\textbf{TopTag systematic validation.}
The constituent split model reduces physics leakage from the nuisance branch across three seeds, and the leakage reduction also appears on a held-out systematic domain. Red circles denote the shared baseline; blue squares denote the split candidate; grey open circles denote the held-out domain check. Filled markers and horizontal bars show mean $\pm$ SD over three seeds for the balanced systematic-domain study. \textbf{a,} Physics-label AUC measured from $z_{\mathrm{nuis}}$; lower values mean less leakage, and the neutral annotation reports the shared-to-split change. The grey dashed open-marker line shows the held-out domain check. \textbf{b,} Top-tagging AUC from $z_{\mathrm{phys}}$. \textbf{c,} Domain-probe accuracy from $z_{\mathrm{nuis}}$. \textbf{d,} Background rejection at 30\% signal efficiency. The grey triangle in panel d is a separate calibration of the same constituent encoder on the reference dataset~\cite{toptag_reference}, not a systematic domain result.}
\label{fig:toptag_transfer}
\end{figure*}

To test whether the same protocol works outside the H4l EveNet feature extraction setting, we repeated it in a TopTag systematic setting (Fig.~\ref{fig:toptag_transfer}). In the first balanced study over systematic shards, the constituent shared baseline reaches physics AUC $0.9522\pm0.0006$ and $z_{\mathrm{nuis}}\rightarrow$physics AUC $0.9347\pm0.0022$. The balanced split model keeps physics AUC at $0.9519\pm0.0001$ and reduces physics leakage from the nuisance channel to $0.6063\pm0.0044$. Background rejection at 30\% signal efficiency is statistically consistent between the two models, with $381.4\pm27.0$ for the baseline and $384.1\pm28.9$ for the split model.

A test on a held-out systematic domain further separates training domains from evaluation domains. Training on nominal, energy scale up/down, cluster energy resolution and cluster position domains, then holding out the bias domain, gives holdout physics AUC 0.9497 for the shared baseline and 0.9499 for the split model. The holdout $z_{\mathrm{nuis}}\rightarrow$physics AUC decreases from 0.9336 to 0.6220. This is a stronger held-out domain check than a probe evaluated inside the training domains, because the held-out domain is not used in the branch routing objective.

The reference dataset calibration gives AUC 0.9710 and background rejection 281 for the small constituent tagger on the Top Quark Tagging Reference Dataset~\cite{toptag_reference}. This shows that the encoder has reasonable tagging capacity before the systematic test is imposed.

TopTag therefore reproduces the large latent routing change without relying on EveNet and preserves it on one held-out systematic domain. That cross-workflow consistency makes a likelihood-facing prediction test meaningful, but it does not answer the test itself.

\subsection{Lower latent leakage does not reliably predict lower likelihood bias}

The central test asks whether a cleaner nuisance branch predicts a smaller analysis-facing bias. The frozen development audit contains 18 model rows from shards 000 and 001: shared, balanced split and frozen-residual models for seeds 41--43 on each shard. Across these rows, the Spearman rank association between nuisance-branch physics AUC and maximum absolute profiled signal-strength bias is 0.036. Six paired transitions meet the prespecified material leakage-reduction threshold, but three of the six do not reduce the maximum absolute bias. The development data therefore identify a proxy-to-endpoint mismatch candidate rather than a monotonic leakage-to-bias relation.

The preregistered shard002 test confirms the discordance for the shared-to-balanced transition (Table~\ref{tab:proxy_failure_confirmation}). Leakage AUC decreases by 0.318--0.336 in all three seeds, while physics AUC changes by less than 0.0011 in magnitude. Nevertheless, the maximum absolute signal-strength bias increases by 0.0090 and 0.0098 in seeds 41 and 42 and decreases by 0.0036 in seed 43. All three pairs pass the material-leakage and task-preservation gates, while two of three are cleaner but not better on the frozen likelihood endpoint. This satisfies the confirmation rule fixed before shard002 access.

\begin{table*}[t]
\centering
\small
\setlength{\tabcolsep}{3pt}
\caption{\textbf{Preregistered shard002 proxy-failure confirmation.} Differences are balanced split minus shared baseline. Negative $\Delta$ leakage indicates a cleaner nuisance branch; negative $\Delta B_{\max}$ indicates a smaller maximum absolute profiled signal-strength bias. The result confirms discordance within this TopTag protocol, not a universal relation across representations or analyses.}
\label{tab:proxy_failure_confirmation}
\begin{tabular}{rrrrrrc}
\toprule
Seed & $\Delta$ leakage AUC & $\Delta$ physics AUC & Shared $B_{\max}$ & Split $B_{\max}$ & $\Delta B_{\max}$ & Bias improved \\
\midrule
41 & -0.3357 & -0.0011 & 0.0134 & 0.0224 & +0.0090 & no \\
42 & -0.3185 & -0.0006 & 0.0048 & 0.0146 & +0.0098 & no \\
43 & -0.3178 & -0.0007 & 0.0091 & 0.0055 & -0.0036 & yes \\
\botrule
\end{tabular}
\end{table*}

Taken together, the development and confirmation results support the title claim at protocol-local strength: reducing aggregate latent readability does not reliably predict a lower frozen likelihood bias. Leakage probes remain useful because they expose a nuisance channel that behaves as a second task classifier and, with controls, diagnose weak routing or collapse. They are not sufficient to certify score-template behaviour; small residual changes in signal-enriched bins can still dominate $B_{\max}$.

\subsection{An analysis-facing residual target gives mixed secondary evidence}

The profile-stress benchmark localizes a feature that the aggregate leakage probe does not encode. In signal-enriched high-score bins, small residual shape differences can mimic a signal excess after profiling. This bin-local sensitivity is consistent with the observed proxy-to-endpoint mismatch, although it does not establish a complete causal mechanism.

As a secondary check, we trained against a residual target derived from frozen score templates rather than mini-batch histograms. The best tested setting preserves classifier performance across six shard/seed settings, with mean AUC loss $-0.0010\pm0.0003$. It reduces the aggregate unmodelled mean $|\mu|$ bias from $0.0099\pm0.0075$ to $0.0056\pm0.0045$ and the aggregate maximum $|\mu|$ bias from $0.0136\pm0.0099$ to $0.0087\pm0.0067$. The setting-level effect remains heterogeneous: mean bias improves in three of six cases and maximum bias in five of six (Table~\ref{tab:toptag_template_stress}). This mixed result shows that an analysis-facing target can alter the tradeoff; it does not identify a validated replacement proxy.

\begin{table*}[t]
\centering
\caption{\textbf{TopTag frozen-template residual stress summary.} The residual target preserves AUC across six shard/seed settings, while bias improvement is not uniform. Here, a setting denotes one event-shard and training-seed combination; shards 000 and 001 are the two development shards, and seeds 41--43 are independent model initializations.}
\label{tab:toptag_template_stress}
\begin{tabular}{lrrrrr}
\toprule
Setting & $\Delta$AUC & \shortstack{Ref. mean\\$|\mu|$} & \shortstack{Residual mean\\$|\mu|$} & \shortstack{Ref. max\\$|\mu|$} & \shortstack{Residual max\\$|\mu|$} \\
\midrule
shard000 seed41 & -0.0007 & 0.0203 & 0.0059 & 0.0237 & 0.0120 \\
shard000 seed42 & -0.0013 & 0.0158 & 0.0024 & 0.0170 & 0.0047 \\
shard000 seed43 & -0.0007 & 0.0015 & 0.0016 & 0.0023 & 0.0019 \\
shard001 seed41 & -0.0015 & 0.0040 & 0.0054 & 0.0059 & 0.0076 \\
shard001 seed42 & -0.0008 & 0.0050 & 0.0040 & 0.0069 & 0.0055 \\
shard001 seed43 & -0.0007 & 0.0125 & 0.0141 & 0.0255 & 0.0204 \\
\botrule
\end{tabular}
\end{table*}

\section{Discussion}
\label{sec:discussion}

Within the studied TopTag protocol, a cleaner nuisance branch does not reliably imply a smaller profiled-likelihood bias. This conclusion is not driven by a negligible intervention: H4l and TopTag both show large leakage reductions with preserved task AUC, and the control suite excludes a weak probe or empty branch as simple explanations. The mismatch instead reflects the different information summarized by the two endpoints. A leakage probe aggregates physics-label readability across a latent branch, whereas $B_{\max}$ can be controlled by localized residual responses in signal-enriched score bins after profiling.

This finding extends, rather than replaces, earlier cautions about nuisance decorrelation. Ghosh and Nachman showed that decorrelating selected generator variations can reduce an estimated theory uncertainty without reducing disagreement with an independent generator~\cite{ghosh2022caution}. INFERNO, uncertainty-aware learning and systematics-aware neural training address a complementary problem by aligning training more directly with an inference objective~\cite{decastro2019inferno,ghosh2021uncertainty,birk2025sannt}. The present contribution lies at the representation-audit layer: it pairs internal nuisance-branch readability with a profiled score-template endpoint and then subjects the relation to a one-shot preregistered event-shard confirmation.

The two-channel structure is an intervention introduced in this study, not a native component of EveNet. H4l uses EveNet only as a fixed single-embedding interface, while TopTag uses a separately trained constituent encoder with the same routing logic. This cross-workflow agreement reduces the chance that the leakage result is peculiar to one EveNet checkpoint, but it does not establish a universal statement about multi-branch representations. The independent shard likewise confirms the frozen shared-to-balanced comparison within one TopTag workflow; it is not cross-dataset or cross-method validation.

The practical implication is an analysis-facing validation sequence rather than a new performance claim. Table~\ref{tab:certification_checklist} records the minimum evidence layers supported by this study. A latent probe remains useful for diagnosing routing, weak separation and branch collapse. Before the representation is used in an analysis claim, however, the score construction, nuisance model, unmodelled stress shifts and likelihood endpoint should be frozen and tested separately. A held-out confirmation then asks whether the proxy-to-endpoint conclusion survives data not used to select the model or thresholds. This sequence does not certify an arbitrary representation; it identifies which claim each layer of evidence can support.

\begin{table*}[t]
\centering
\caption{\textbf{Minimum analysis-facing validation checklist suggested by the study.} The rows form a sequence: passing an earlier diagnostic does not replace a later one.}
\label{tab:certification_checklist}
\begin{tabular}{@{}L{0.18\linewidth}L{0.31\linewidth}L{0.41\linewidth}@{}}
\toprule
Layer & Required question & Evidence required before a robustness claim \\
\midrule
Task preservation & Does the intervention retain the intended physics discrimination? & Paired task AUC or operating-point performance under identical event splits and seeds. \\
Representation audit & Is nuisance-branch physics readability reduced without a weak probe or collapsed branch? & Reinitialized probes, positive and shuffled controls, and representation-rank diagnostics. \\
Analysis-facing stress & Does the actual score-template likelihood remain stable under a prespecified unmodelled shift? & Profiled parameter bias or interval behaviour from a frozen likelihood and template construction. \\
Held-out confirmation & Does the conclusion survive data not used to select the metric, model or thresholds? & Preregistered event shard within the same workflow, fixed decision rules, complete provenance and no post-access sweep. \\
\botrule
\end{tabular}
\end{table*}

Several boundaries remain. The H4l domains are synthetic visible controls rather than official CMS systematic variations, its observed-data overlay is exploratory and not deduplicated across streams, and its nominal $m_{4\ell}$ template is more precise than either learned-score template. The TopTag likelihood is a reproducible HistFactory-style stress benchmark on selected public shards rather than an ATLAS measurement. The frozen-residual target improves aggregate bias summaries but not every setting, so it is secondary evidence that analysis-aligned objectives can change the tradeoff, not a validated replacement proxy.

A broader claim would require applying the same audit to independently published decorrelation or nuisance-aware methods and to analysis-specific nuisance models. That extension is intentionally outside the present scope. The current result instead supplies a falsifiable case study and a reproducible validation rule: latent-space metrics can screen representations, but likelihood-facing behaviour must be measured rather than inferred from representation cleanliness.

\section{Conclusion}
\label{sec:conclusion}

This controlled audit finds a protocol-local mismatch between a representation diagnostic and a likelihood-facing endpoint. Substantial latent-leakage reductions preserve task AUC, yet they do not reliably predict a smaller maximum absolute profiled signal-strength bias. The one-shot shard confirmation strengthens this conclusion for the frozen TopTag comparison without turning three seeds into a population-level significance claim. The result neither generalizes the mismatch to all latent probes, representations or nuisance models nor supplies a replacement proxy. It instead fixes the role of the tested diagnostic: use latent readability to inspect representation routing, then test any inference-robustness claim with a prespecified score-template likelihood and held-out data. Extending the conclusion beyond this protocol will require independent nuisance-aware methods and analysis-specific likelihood models, not additional tuning on the confirmation shard.

\appendix
\section{Supplementary validation and reproducibility}
\label{app:supplementary_validation}

\subsection{Exploratory observed-data workflow check}

This appendix records a label-free observed-data overlay for the H4l workflow. It serves only as an exploratory workflow sanity check.

\begin{figure*}[t]
\centering
\includegraphics[width=0.80\textwidth]{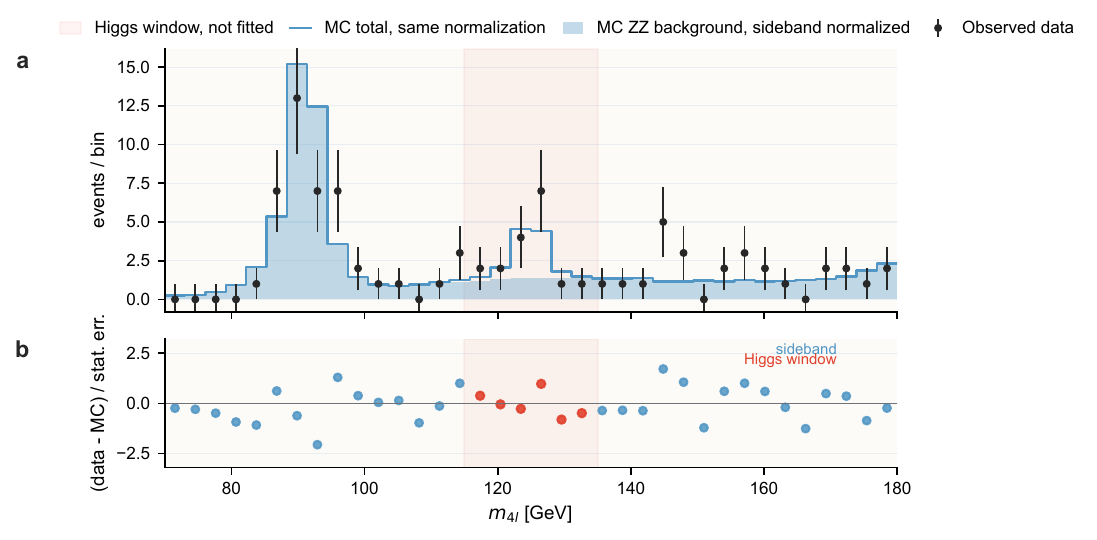}
\caption{\textbf{Observed data H4l workflow sanity check.}
\textbf{a,} Observed selected events are overlaid with simulated MC templates in $m_{4\ell}$ after sideband normalization. Black points show observed counts, the filled blue histogram shows MC $ZZ$ background and the blue step shows MC total. The pale red band marks the Higgs window used for context but not fitted. \textbf{b,} Residuals, $(\mathrm{data}-\mathrm{MC})/\mathrm{stat.\ err.}$, are shown on the same $m_{4\ell}$ axis. Blue and red labels identify sideband and Higgs-window bins, respectively. The non-deduplicated overlay is not a signal strength measurement, discovery fit or profiled likelihood validation.}
\label{fig:h4l_boundary}
\end{figure*}

\clearpage
\subsection{H4l likelihood scans}

\begin{figure*}[t]
\centering
\includegraphics[width=0.80\textwidth]{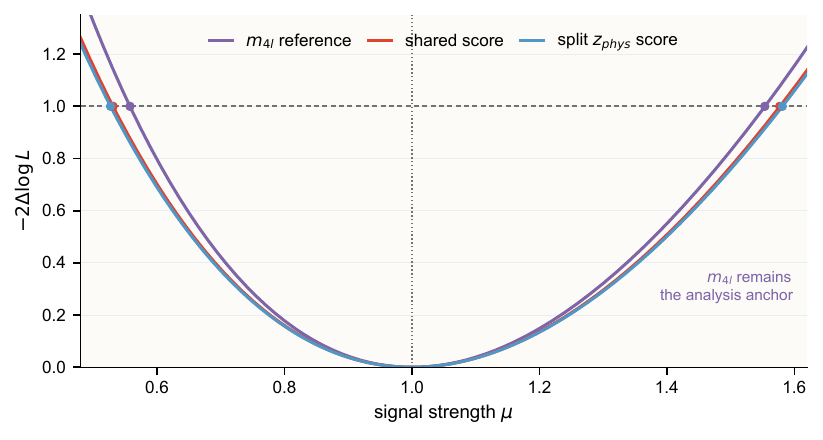}
\caption{\textbf{Supporting H4l nominal template scans.}
Lightweight binned MC/Asimov scans compare the conventional $m_{4\ell}$ template with shared and split learned-score templates near the one-sigma threshold. Purple, red and blue curves show the $m_{4\ell}$ reference, shared-score template and split $z_{\mathrm{phys}}$ score template, respectively. The horizontal dashed line marks $-2\Delta\log L=1$, and filled points mark approximate one-sigma crossings. The vertical dotted line marks $\mu=1$. The curves use nominal validation MC templates and do not profile systematic nuisances.}
\label{fig:appendix_h4l_likelihood}
\end{figure*}

The H4l nominal template MC/Asimov scans provide the corresponding check at the analysis boundary for the main workflow (Fig.~\ref{fig:appendix_h4l_likelihood}). The finely binned $m_{4\ell}$ template remains the strongest anchor for likelihood construction, while the shared and split learned scores remain auxiliary representation diagnostics.

These studies are useful sanity checks but do not imply profiled likelihood reduction in a real experiment model. They motivate the boundary used in the main text: the present work is a representation diagnostic, not a substitute for likelihood construction, nuisance profiling or official systematic propagation.

\clearpage
\subsection{Controlled EveNet raw-shard shift studies}
\label{app:raw_shard_studies}

The main analysis uses the CMS H4l open-data-derived EveNet workflow as the primary H4l evidence and treats the earlier EveNet raw-shard studies as mechanism and boundary evidence. These raw-shard studies are retained because they explicitly test matched-event response under controlled jet-energy-scale (JES), missing-transverse-momentum (MET) and combined JES+MET perturbations, but they do not by themselves establish detector-nuisance learning.

In the corrected 20-raw evaluation, the best candidate reduced the worst-case latent shift ratio from 1.7974 for the shared-latent baseline to 0.7399, and reduced the worst-case observable-view shift ratio from 1.9417 to 0.8891. The paired repeated-probe physics-readability difference was $-0.0291 \pm 0.0173$. This result supports the branch-stability mechanism, while the readability cost and proxy-based nuisance definition keep the interpretation bounded.

As an intermediate multi-raw check, we also evaluated the same candidate family on five EveNet raw shards. This replication used the same fixed EveNet teacher, the same physics-versus-nuisance readout family, the same JES/MET systematic-shift protocol and the same repeated-probe readability check. In the five-raw replication, the candidate reduced the worst-case latent shift ratio from 1.6150 to 1.0375. The worst-case observable-view shift ratio decreased from 1.4468 to 1.1644. The paired repeated-probe difference was $-0.0024 \pm 0.0213$, placing the readability change within repeated-probe uncertainty.

\subsection{Response-aware joint-objective boundary test}
\label{app:response_aware_boundary}

We tested a stronger response-aware readout objective to evaluate whether the stability-readability tradeoff could be removed by explicitly allocating systematic response to a separate branch. The objective used a response-context readout with $z_{\mathrm{phys}}$, $z_{\mathrm{nuis}}$ and $z_{\mathrm{resp}}$, matched consistency on $z_{\mathrm{phys}}$, paired-contrastive geometry, nuisance suppression on $z_{\mathrm{phys}}$, and a response-allocation loss on the response branch. The hard-negative pair-triplet term was disabled in the final run because its backward pass was numerically unstable in this setting; the paired-contrastive term was retained.

Table~\ref{tab:response_aware_boundary} summarizes the two 20-raw response-aware variants. A candidate was treated as passing the joint gate only if it simultaneously satisfied the physics-readability floor, nuisance-leakage ceiling, latent-stability ceiling, observable-stability ceiling and event-retrieval floor. Neither variant passed this joint criterion. The balanced variant retained acceptable physics readability and nuisance suppression and had favourable latent and observable stability, but its mean same-class event-retrieval top-1 accuracy was below the event-preservation threshold. The stronger suppression variant further reduced nuisance readability from $z_{\mathrm{phys}}$, but worsened latent stability and event retrieval. These results support the interpretation that response allocation can move individual axes of the tradeoff without producing a Pareto-dominant physics branch.

\begin{table*}[t]
\centering
\caption{\textbf{Response-aware joint-objective boundary test at 20 raw shards.} Arrows indicate preferred directions. Neither tested variant passes the joint gate.}
\label{tab:response_aware_boundary}
\resizebox{\textwidth}{!}{%
\begin{tabular}{lrrrrrrrr}
\toprule
Model & $z_{\mathrm{phys}}\!\to$phys $\uparrow$ & $z_{\mathrm{phys}}\!\to$nuis $\downarrow$ & latent worst $\downarrow$ & observable worst $\downarrow$ & top-1 $\uparrow$ & yield drift $\downarrow$ & $\mu$ drift $\downarrow$ & gate \\
\midrule
joint\_balanced\_safe & 0.3579 & 0.5451 & 0.6580 & 0.8405 & 0.2255 & 0.1067 & 0.0308 & fail \\
joint\_suppress\_safe & 0.3637 & 0.5359 & 0.8393 & 0.6274 & 0.1663 & 0.1360 & 0.0312 & fail \\
\botrule
\end{tabular}
}
\end{table*}

The branch-combination diagnostic was informative. For the balanced variant, the mean same-class top-1 retrieval of $z_{\mathrm{phys}}$ alone was 0.2100, 0.2456 and 0.2209 for JES, MET and combined JES+MET, whereas the concatenated $z_{\mathrm{phys}}+z_{\mathrm{resp}}$ representation reached 0.3354, 0.3612 and 0.4096. For the stronger suppression variant, the corresponding $z_{\mathrm{phys}}$ retrieval values were 0.1061, 0.2597 and 0.1331, while the concatenated representation reached 0.3800, 0.3521 and 0.3590. Thus event- and shift-response information remained recoverable from the combined representation, but was not preserved inside $z_{\mathrm{phys}}$ alone.

\subsection{Frozen proxy-to-likelihood audit and confirmation protocol}
\label{app:proxy_likelihood_protocol}

The development audit used only exact checkpoints and profile-stress outputs already obtained on TopTag event shards 000 and 001. For each shard and each of seeds 41--43, it evaluated the shared baseline, balanced split and frozen-residual candidate. Three newly initialized probes with seeds 8201--8203 were fitted to each frozen nuisance branch. Across the resulting 18 model rows, the Spearman rank association between mean nuisance-branch physics AUC and maximum absolute profiled signal-strength bias was 0.0361. Six paired transitions reduced leakage AUC by at least 0.05, and three of these six did not reduce the likelihood bias. The audit was retrospective and was used only to preregister a separate confirmation.

Before shard002 was accessed, the confirmation was restricted to the shared-to-balanced transition for seeds 41--43. The training epochs, batch size, learning rate, event cap, constituent cap, probe seeds, template binning, profile settings and decision thresholds were fixed. The preregistered configuration and controller script were frozen and recorded with file-level SHA-256 checksums in the reviewer archive manifest. The selected nominal, energy-scale-up, energy-scale-down, resolution, position and bias files were frozen by URI, byte size and Adler-32 checksum. A no-access preflight confirmed that neither raw nor cached shard002 files were present before execution.

The confirmation produced six finite model rows and three complete paired contrasts. Leakage AUC differences for seeds 41, 42 and 43 were $-0.33565$, $-0.31849$ and $-0.31785$, while physics AUC differences were $-0.00107$, $-0.00060$ and $-0.00075$. The corresponding differences in maximum absolute profiled signal-strength bias were $+0.00896$, $+0.00982$ and $-0.00365$. Thus all pairs met the material-leakage and task-preservation gates, while two of three were cleaner but not better. No post-access seed, architecture, probe, likelihood, threshold or proxy sweep was performed. This one-shot result confirms the proxy--endpoint discordance on an independently accessed event shard within the frozen TopTag workflow, but it does not identify a replacement proxy or establish cross-method generality.

\subsection{Training and probe settings}
\label{app:training_settings}

Table~\ref{tab:training_settings} records the run settings needed to reproduce the manuscript-level conclusions from the public scripts, source summaries and the pending tensor/checkpoint archive. Probe heads are trained after the representation model is frozen. The table is not a substitute for the run manifests; rather, it makes the settings that define the reported comparisons visible in the paper.

The H4l downstream network uses a 128-unit layer-normalised GELU trunk, a 64-dimensional physics branch and a 16-dimensional nuisance branch for the selected setting. The TopTag network maps four constituent features through two 64-unit pointwise GELU layers, concatenates masked mean and maximum pooling with a layer-normalised high-level feature branch, and uses a 128-unit shared trunk, a 64-dimensional physics branch and a 32-dimensional nuisance branch. Both workflows use AdamW with weight decay $10^{-4}$ and training-partition standardisation; class imbalance enters through the positive-class weight in the binary cross entropy. Linear post hoc probes are trained only after the representation checkpoint is fixed. Reported ``$\pm$'' values are sample standard deviations across seeds unless stated otherwise.

\begin{table*}[t]
\centering
\caption{\textbf{Training and probe settings used for the manuscript comparisons.} Values are taken from the run scripts and the reports used to generate the figure-source data.}
\label{tab:training_settings}
\scriptsize
\setlength{\tabcolsep}{2pt}
\begin{tabular}{@{}L{0.18\linewidth}L{0.23\linewidth}L{0.22\linewidth}L{0.20\linewidth}L{0.11\linewidth}@{}}
\toprule
Study & Inputs and split & Model settings & Optimisation and probes & Source \\
\midrule
H4l branch routing & 218618 CMS H4l MC candidates; mixed visible-domain copies; validation ratio 0.2 & Fixed EveNet embeddings; split physics and nuisance branches; nuisance latent dim 16; adversarial/domain weight 0.5; orthogonality weight 0.5 & 5 training epochs; batch size 8192; embedding batch size 4096; learning rate $10^{-3}$; post hoc probes for 20 epochs & \url{scripts/e73_cms_h4l_evenet_domain_design_repeat.py}; E73 reports \\
TopTag systematic audit & ATLAS TopTag record 80030; nominal, esup, esdown, cer and cpos training domains; bias held out for the held-out audit; validation ratio 0.2 & Constituent split network; hidden dim 128; latent dim 64; nuisance latent dim 32; nuisance weight 2.0; physics-adversarial weight 0.5; max 80 constituents & 6 training epochs; batch size 4096; learning rate $8\times10^{-4}$; orthogonality weight 0.25; post hoc probes for 8 epochs & \url{scripts/e77_toptag_heldout_systematic_audit.py}; E77 reports \\
TopTag metric-to-likelihood audit & Development shards 000/001 and preregistered shard002; seeds 41--43; 20 score bins & Frozen shared/balanced checkpoints; E102 also audits frozen-residual checkpoints; profiled energy-scale nuisance and unmodelled-shift pseudo data & Three post hoc probe seeds; material leakage threshold $-0.05$; physics-AUC floor $\Delta\mathrm{AUC}\geq-0.01$; no post-access sweep & E102/E103 scripts and reports \\
TopTag frozen-template residual target & Same balanced TopTag shards; frozen checkpoint from the score-template export; residual target applied to high-score template bins & Same constituent split candidate as the TopTag audit; nuisance latent dim 32; residual target from fixed score-template residuals & 6 training epochs; batch size 4096; learning rate $8\times10^{-4}$; residual weight 0.03; residual minimum score 0.75; residual temperature 0.04 & \url{scripts/e91_toptag_frozen_template_residual_target.py}; E91 reports \\
Controlled EveNet raw-shard checks & Twenty raw shards for the corrected boundary evaluation and five raw shards for replication; matched JES, MET and combined JES+MET perturbations & Same physics-versus-nuisance readout family as the controlled-shift studies; response-aware variants reported in Appendix~\ref{app:response_aware_boundary} & Candidate selection and repeated-probe settings follow the archived run manifests and reports; these studies provide mechanism and boundary evidence rather than the main H4l claim & E52/E45 and response-aware reports \\
\botrule
\end{tabular}
\end{table*}

\clearpage
\section*{Data and code availability}

\paragraph{Public data.}
The H4l workflow uses CMS open-data records 12361--12368 linked from record 12360~\cite{cms_h4l_opendata}. The TopTag workflow uses record 80030~\cite{atlas_toptag80030}, and its capacity calibration uses the TopTag reference dataset~\cite{toptag_reference}. The fixed EveNet feature extractor uses the public checkpoint cited in this manuscript~\cite{hsu2026evenet}.

\paragraph{Code and figure source data.}
Scripts for preprocessing, branch training, probe fitting, control checks and figure generation are available at \url{https://github.com/anonymous375335/event-representation-reliability-review}. Figure-source comma-separated value files are stored under \texttt{figures/}. The TopTag template benchmark is under \texttt{benchmarks/toptag\_pyhf/}, with selected-shard manifests, example workspaces, wrapper scripts and E79--E91 summary files.

\paragraph{Large artifacts.}
Large derived tensors and checkpoints are not stored in Git. The anonymous review artifact archive (version 2) is retained as a hash-verified review artifact and can be supplied to the editors or reviewers through an agreed transfer route on request. The 251906507-byte ZIP has SHA-256 \texttt{\seqsplit{7df035f1fdb9c0458e753c68969452ce1f216c9d4e0120d6f29d58007ed48f10}}; its 172-file payload manifest and original-binary availability table are tracked in the reviewer repository. The archive contains the exact H4l tensor and EveNet checkpoint, the H4l headline runs and controls, preserved development E79 runs, E79/E91d reports and summaries, and the complete E102 audit and E103 confirmation evidence. All three shard002 E79/E81 seed runs are preserved as originals. Development settings whose original GPU run directory was not synchronized remain marked as regeneration-only; no replacement checkpoint is represented as an archived original. A public archival DOI will be added to the accepted version.

The public-input preprocessing and figure-source workflows can be inspected from the repository. The review archive supplies the derived H4l inputs and the preserved original TopTag binaries, while the public-input manifests and frozen scripts define regeneration for the remaining TopTag settings. Table~\ref{tab:repro_checklist} lists the required artifacts.

\begin{table*}[t]
\centering
\caption{\textbf{Reviewer reproducibility checklist.} Large inputs are external to git; scripts and source summaries are kept in the anonymous review package.}
\label{tab:repro_checklist}
\begin{tabular}{@{}L{0.19\linewidth}L{0.34\linewidth}L{0.37\linewidth}@{}}
\toprule
Component & Artifact visible to reviewers & Purpose \\
\midrule
Public inputs & CMS records 12360--12368; ATLAS record 80030; TopTag reference dataset & Defines the public data sources used by the two workflows \\
Derived tensors and checkpoints & Anonymous review archive with archive and file-level SHA-256 manifests; public DOI after acceptance & Supplies exact H4l derived inputs and preserved TopTag binaries, with regeneration-only settings identified explicitly \\
H4l reliability scripts & H4l preprocessing, EveNet embedding, branch routing and control scripts & Reproduces the main H4l AUC, leakage, domain readout and control results \\
TopTag reliability scripts & TopTag systematic, held-out domain, E102 audit and E103 confirmation scripts & Reproduces the TopTag workflow, frozen metric-to-likelihood audit and preregistered event-shard confirmation within that workflow \\
Figure package & \texttt{figures/} source CSV files and figure generation script & Regenerates the manuscript figures and audit tables \\
\botrule
\end{tabular}
\end{table*}

\section*{Statements and Declarations}

\paragraph{Competing interests.}
The author declares no competing interests.

\paragraph{Funding.}
No funding was received for this work.

\paragraph{Author contributions.}
Tong Pan designed the study, implemented the workflows, analysed the results and wrote the manuscript.

\paragraph{Data and code availability.}
Data and code availability are described in the preceding section.

\paragraph{Ethics approval.}
Not applicable.

\end{document}